\documentclass[%
 reprint,
 amsmath,amssymb,
 aps,
]{revtex4-2}

\usepackage{graphicx}% Include figure files
\usepackage{dcolumn}% Align table columns on decimal point
\usepackage{bm}% bold math
\usepackage[utf8]{inputenc}
\usepackage[T1]{fontenc}
\usepackage{mathptmx}
\usepackage[dvipsnames]{xcolor}
\usepackage{multirow}

\begin{document}

\preprint{AIP/123-QED}

\title{Chern Invariants of Topological Continua; a Self-Consistent Nonlocal Hydrodynamic Model}
% Force line breaks with \\

\author{S. Pakniyat}
\email{pakniyat@uwm.edu}
\affiliation{Department of Electrical Engineering, University of Wisconsin-Milwaukee, Milwaukee, Wisconsin 53211, USA}%

\author{S. Ali Hassani Gangaraj}
\email{ali.gangaraj@gmail.com}
\affiliation{School of Electrical and Computer Engineering, University of Wisconsin-Madison, Madison, WI 53706, USA}%

\author{G.W. Hanson}
\email{george@uwm.edu}
\affiliation{Department of Electrical Engineering, University of Wisconsin-Milwaukee, Milwaukee, Wisconsin 53211, USA}

\date{\today}
            
\begin{abstract}
 Abstract- Topological systems are characterized by integer Chern invarients. In a continuous
photonic system characterized by a local Drude model, the material response is ill-behaved at large wavenumbers,
leading to non-integer Chern invarients and ambiguity in the existence of topological edge modes. This
problem has been solved previously by introducing an ad hoc material model including a spatial cutoff material’s
wavenumber, which leads to a finite Brillouin zone and integer invarients. In this work, we calculate Chern
numbers in magnetized continuous plasma systems by considering the effect of nonlocality using a hydrodynamic
Drude model. Then, we argue that
this model presents several advantages compared to the previous models, e.g. introducing
physical response at large wave numbers and integer Chern invarients with sum to zero without the need for an
interpolated material response. Therefore, the hydrodynamic model forms a complete and self-consistent model,
which resolves the Chern number issues in topological photonic continua.
\end{abstract}

\maketitle

\section{\label{sec:level1}Introduction}

\bigskip Topological insulators, including artificial periodic structures and continuous materials \cite{khanikaev2013photonic, hasan2010colloquium,davoyan2013theory,lu2014topological, jin2016topological,rechtsman2013photonic,tsakmakidis2017breaking,gao2016photonic, shastri2021nonreciprocal,gangaraj2016effects,gangaraj2019unidirectional,pakniyat2020non},      and recently topological polaritonics systems  \cite{karzig2015topological,li2020experimental,guddala2021topological} have been broadly studied in the last two decades. Magnetized plasma systems are classified as Chern-type insulators with broken time reversal symmetry. They are
characterized by a topological index known as the Chern number \cite{sil2015chern, gangaraj2017berry}. This number cannot change except when the underlying momentum space topology of the bulk bands is changed. For instance, this occurs when a bandgap opens or closes. One of the most important features of topological materials is that they support unidirectional surface plasmon polaritons (SPPs) with unique properties. In topological photonic insulators with broken time reversal symmetry (nonreciprocal), the relevant topological invariant is the gap Chern number, i.e., the sum of the Chern numbers of all bulk modes below the bandgap. The edge states connect different energy levels of the bulk modes. If the edge state has a nonreciprocal response within the bandgap of the
nontrivial bulk modes, it is a wave protected from
back-scattering and diffraction. In other words, it is unaffected by smooth deformations in the surface that preserve topology (note that surface geometry may include sharp features). The bulk-edge correspondence principle links the Chern invariants of two topological insulators having a common bandgap with the number of unidirectional SPP modes that exist at the interface of the two materials \cite{silveirinha2016bulk,sil2019proof, tauber2020anomalous}. While this principle works well for topological photonic insulators based on periodic structures, subtle issues arise in the case of topological photonic continua due the absence of intrinsic periodicity. Reference \cite{gangaraj2020physical}, studied two general classes of the bulk-edge correspondence principle violations for continuous topological photonic materials: (i) inconsistency between the gap chern number and the number of edge states and (ii) incomplete gap coverage by the edge state line. As further discussed in \cite{gangaraj2020physical}, these violations are associated with the asymptotic behavior of the surface modes for large wave numbers. In addition, it has been shown that although the above-mentioned violations can be restored by adding hydrodynamic nonlocality, the correspondence principle is physically violated for practical purposes, even with zero intrinsic bulk losses, due to Landau damping or nonlocality-induced radiation leakage. The present work focuses on the bulk modes in continuous photonic media and the issues of noninteger Chern numbers and their nonzero sum. In \cite{sil2015chern} it has been shown that the former can be resolved by adding a spatial cut-off wavenumber and the latter can be addressed by interpolating the interfaced material models. However, in this work by comparing all different nonlocal models, their effects on Chern number, bulk bands and their pros and cons, we show that the hydrodynamic nonlocal model can resolve the two issues associated with Chern numbers all at once.

In Refs. \cite{raghu2008analogs,haldane2008possible} a method for Chern number calculation in periodic photonic crystals has been introduced by Raghu and Haldane. Then, Silveirinha has developed this method for anisotropic continua \cite{sil2015chern}. He found that the Chern
numbers in continuous materials are integer invariants subject to
considering spatially dispersive material models. He has introduced an ad hoc nonlocal material model having a large spatial cutoff wavenumber. Through the paper, this model is called the spatial cutoff model. By this assumption, the Hamiltonian becomes well-behaved at large momentum, unlike in the local material model. As a result, integer Chern invariants of $\{+1,-2\}$ are obtained, respectively for high and low frequency bands of the TM bulk modes, which does not add up to zero as required. To solve this problem, it was suggested to apply an interpolate material model which represents a continuous transition from an isotropic plasma to a gyrotropic plasma medium. Hereupon, a new frequency band appears at very low frequencies, whose Chern number is +1. Therefore, the appearance of this new low frequency band resolved the issue of nonzero summation of Chern numbers, but at the expense of a complicated permittivity interpolation of interfaced materials. Although this is a clever way to fix the deficiencies of the model, this ad hoc spatial cutoff model does not provide a self-contained description of plasmonic materials. In this regard, we propose to solve the non-integer Chern invariants of the plasma continua via solving the hydrodynamic equation in the magnetized plasma for continuous photonic topological platforms. We evaluate the effect of nonlocality due to the pressure (a manifestation of non-locality) on topology of the bulk modes and the associated Chern numbers in magnetized plasma systems. For the hydrodynamic model, we calculate the band Chern numbers and obtain integer invariants of $\pm1$, which guarantees topological behavior by considering a realistic nonlocal material model.

Nonlocality plays an important role in the unidirectional nature
of the surface plasmon polaritons in topological systems. In ref. \cite{buddhiraju2020absence}, it has been
argued that by modeling a gyrotropic plasma using the hydrodynamic Drude model, a truly
unidirectional SPP will not exist at the
interface of the dielectric and magnetized plasma media below the plasma frequency due to the effect of nonlocality (i.e., there will always be a backward mode, although perhaps only existing at large wavenumber, which may be relatively unimportant from a practical standpoint). It has been demonstrated that the surface waves have nonreciprocal
bi-directional propagation behavior, which is in contradiction with purely unidirectional propagation behavior predicted by applying a simple Drude model. But in Refs. \cite{gangaraj2019truly, monticone2020truly}, it has been clarified that the nonlocality does
not affect a class of unidirectional SPPs that exists at the interface of
opaque and magnetized plasma media above the plasma frequency.
Recently, the unidirectional properties of this class of the surface waves have been experimentally verified in plasma systems \cite{tunable,pakniyat2020indium}. 
Following these studies, we obtain integer Chern invarients to formally validate the existence of topological
unidirectional SPPs by considering realistic conditions using the nonlocal hydrodynamic model.  

In the following, first we derive a dielectric tensor to characterize a
gyrotropic medium using the hydrodynamic model. Then, we evaluate the bulk mode
properties in a magnetized nonlocal plasma region. Finally, we calculate Chern numbers for the hydrodynamic model and compare the results with local and spatial cutoff models.  

\section{\label{sec:level3}Dielectric tensor of the hydrodynamic model}

 Consider a plasma medium consisting of $n_{e}$ free electrons per volume with
the effective mass of $m^{\ast }$, electron charge $e$, and mobility $\mu $. A static
magnetic field bias $\mathbf{B}=B_{0}\mathbf{\hat{b}}_{c}$ is applied in
the plasma region, where $B_{0}$ is the magnetic field intensity and $%
\mathbf{\hat{b}}_{c}$ is a unit vector along the magnetic field vector. In
the hydrodynamic model, the equation of motion of the particles is \cite{raza2015nonlocal}
\begin{equation}
\frac{d\mathbf{v}}{dt}+\gamma \mathbf{v+}\left( \mathbf{v\cdot \nabla }%
\right) \mathbf{v}=\frac{e}{m^{\ast }}\left( \mathbf{E(r},t)+(\mathbf{v}%
\times \mathbf{B})\right) -\beta ^{2}\frac{\nabla n(\mathbf{r},t)}{n}
\end{equation}
There are three forces acting on the free electrons, $e\mathbf{E}$ arising
from the electric field of the wave and $e(\mathbf{v}\times \mathbf{B})$
arising from the motion of the electrons with the average velocity of $%
\mathbf{v}$ through the static magnetic field $\mathbf{B}$ (here we ignore the small self-consistent time-varying magnetic field). The last
term represents pressure, where $\beta $ is a nonlocal parameter proportional to the Fermi velocity $\upsilon _{F}
$ in the semiconductor; $\beta ^{2}=\upsilon
_{F}^{2}(3/5\omega +1/3i\gamma)/(\omega +i\gamma)$ \cite{halevi1995hydrodynamic}. In the local model, the induced
charge distribution is assumed to be confined to the boundary of the plasma region by a Dirac
delta function. However, in the hydrodynamic model, the induced charge
density spreads into the bulk plasma region with charge distribution
depth of $\delta =\beta /\omega _{p}$ which is a function of the nonlocal parameter $%
\beta $ \cite{ciraci2013hydrodynamic}.

By linearizing the equation of motion, and considering the continuity
equation $\partial _{t}n=-\mathbf{\nabla .(}n\mathbf{v)}$ and  $\mathbf{J}_{c}=-n_{e}e\mathbf{\nu }$, the induced current
equation is given by \cite{raza2015nonlocal}
\begin{equation}
\beta ^{2}\nabla \left( \nabla \cdot \mathbf{J}_{c}\right) +\omega (\omega 
\mathbf{+}i\gamma )\mathbf{J}_{c}=i\omega \left( \omega _{p}^{\ast
2}\varepsilon _{0}\varepsilon _{\infty }\mathbf{E(r,\omega )-}\omega _{c}%
\mathbf{J}_{c}\times \mathbf{\hat{b}}_{c}\right)
\end{equation}%
where $\omega _{c}=-eB_{0}/m^{\ast }$, $\omega _{p}^{\ast }=\omega
_{p}/\sqrt{\varepsilon _{\infty }}$ ($\omega _{p}=\sqrt{n_{e}e^{2}/m^{\ast
}\varepsilon _{0}}$) and $\gamma =-e/\mu m^{\ast }$ are the
cyclotron, reduced plasma, and collision frequencies, respectively, $%
\varepsilon _{\infty }$ is the high frequency dielectric constant and $%
\varepsilon _{0}$ is the free-space permittivity. By the spatial
Fourier transform and considering $\mathbf{J}_{c}\mathbf{(}k,\omega \mathbf{)%
}=\mathbf{\bar{\sigma}(}k,\omega \mathbf{)}\cdot \mathbf{E(}k,\omega \mathbf{%
)}$, the conductivity tensor is governed by
\begin{equation}
\mathbf{\bar{\sigma}(}k\mathbf{,}\omega \mathbf{)}=i\omega \varepsilon
_{0}\varepsilon _{\infty }X\left( \mathbf{\bar{I}}-iY\mathbf{\hat{b}}%
_{c}\times \mathbf{\bar{I}}\right) ^{-1}
\end{equation}%
where $X=\omega _{p}^{\ast 2}/(\omega (\omega \mathbf{+}i\gamma
)-\beta ^{2}k^{2})$ and $Y=\omega \omega _{c}/(\omega \left( \omega 
\mathbf{+}i\gamma \right) -\beta ^{2}k^{2})$. By taking into consideration that the inverse of a
tensor in the form of $\mathbf{\bar{C}=}$ $\lambda \mathbf{\bar{I}+c}\times 
\mathbf{\bar{I}}$ is $\mathbf{\bar{C}}^{-1}=adj(\mathbf{\bar{C}})/\left\vert 
\mathbf{\bar{C}}\right\vert$, where $\left\vert \mathbf{\bar{C}}\right\vert
=\lambda (\lambda ^{2}+c^{2})$ and $adj(\mathbf{\bar{C}})=\lambda (\lambda 
\mathbf{\bar{I}-c}\times \mathbf{\bar{I}})+\mathbf{cc}$, where
$\mathbf{cc}$ stands for complex conjugate of the first bracket
term, we obtain the dielectric tensor as
\begin{figure*} 
	\includegraphics[width=1.7\columnwidth]{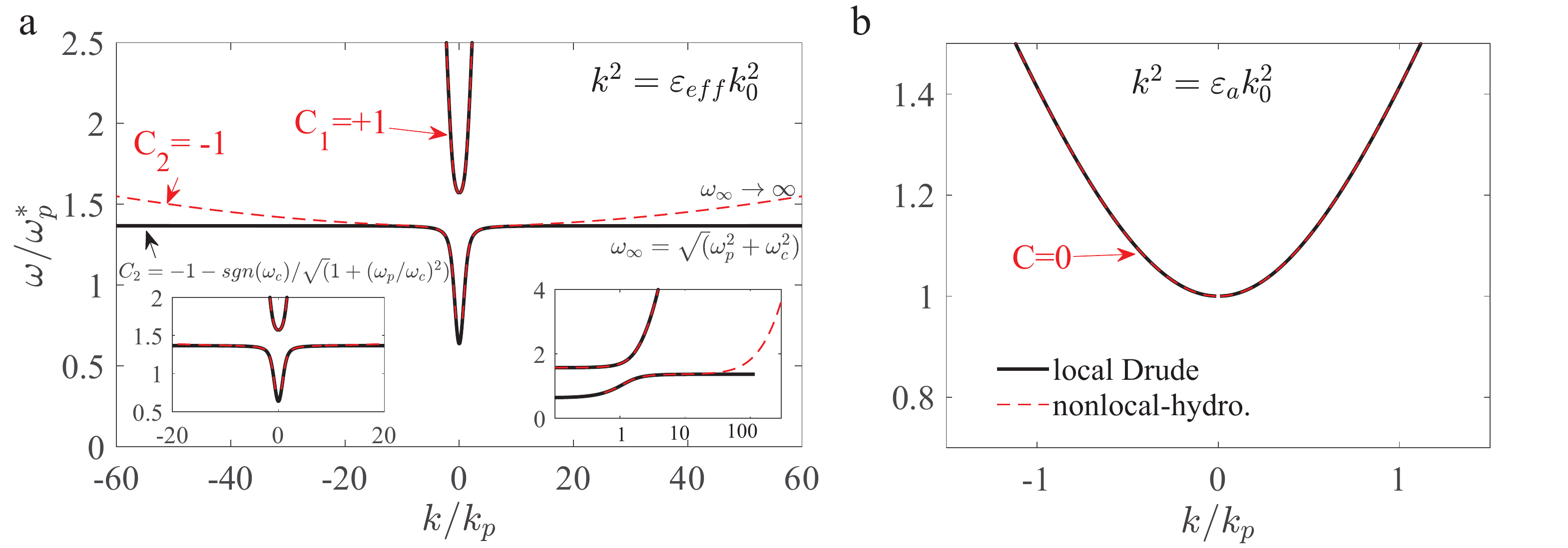}
  \caption{Dispersion bands and associated Chern numbers of (a) nontrivial bulk modes and (b) trivial modes using the local Drude and nonlocal hydrodynamic Drude models; $k_p=\omega_p/c_0$, where $c_0$ is the speed of light in free space. The magnetized plasma region is modeled by (\ref{tens}), using the parameters $n_{e}$ $=$ $3.6\times 10^{21}($m$^{-3})$, $%
\varepsilon _{\infty }=15.68$, $m^{\ast }=0.0175$m$_{0}$, $B_{0}=0.6$T, $\mu
=\infty $, corresponding to $\omega _{p}^{\ast }=2\pi (1.03$THz$)$, $\omega _{c}/\omega
_{p}=0.23$ and $\gamma =0$, related to the InSb crystal at low temperature \cite{liang2021temperature}, and the nonlocality parameter of $\beta=0.77\times 10^6$ m/s.
	}
		 \label{disp}
\end{figure*}

\begin{equation}
\begin{aligned}
\mathbf{\bar{\varepsilon}(}k\mathbf{,}\omega \mathbf{)} &= \varepsilon
_{\infty }\left( \mathbf{\bar{I}+}i\frac{1}{\omega \varepsilon _{0}}\mathbf{%
\bar{\sigma}(}k\mathbf{,}\omega \mathbf{)}\right) \\
&=\varepsilon _{\text{t,nl}
}(\mathbf{\bar{I}-\hat{b}}_{c}\mathbf{\hat{b}}_{c})+i\varepsilon _{\text{g,nl%
}}(\mathbf{\hat{b}}_{c}\times \mathbf{\bar{I}})+\varepsilon _{\text{a,nl}}%
\mathbf{\hat{b}}_{c}\mathbf{\hat{b}}_{c}
\label{tens}
\end{aligned}
\end{equation}%
where the permittivity elements are defined as 
\begin{eqnarray}
\varepsilon _{\text{a,nl}}(k,\omega ) &=&\varepsilon _{\infty }\left(
1-X\right) =\varepsilon _{\infty }-\frac{\omega _{p}^{2}}{\Omega_{k}}\text{ } \\
\varepsilon _{\text{t,nl}}(k,\omega ) &=&\varepsilon _{\infty }\left( 1%
\mathbf{-}\frac{X}{1-Y^{2}}\right) =\varepsilon _{\infty }\mathbf{-}\frac{%
\omega _{p}^{2}\Omega_k }{\Omega_k ^{2}-\left( \omega \omega _{c}\right) ^{2}} \\
\varepsilon _{\text{g,nl}}(k,\omega ) &=&\varepsilon _{\infty }\frac{-YX}{%
1-Y^{2}}=\frac{-\omega \omega _{c}\omega _{p}^{2}}{\Omega_k ^{2}-\left( \omega \omega _{c}\right) ^{2}}
\end{eqnarray}%
with $\Omega_k=\omega (\omega \mathbf{+}i\gamma )-\beta ^{2}k^{2}$. By assuming $\varepsilon _{\infty }=1$
and\ $\gamma =0$ and defining a nonlocal factor as $\chi =1/(1-k^{2}/k_{\text{m}}^{2})$, where $k_{\text{m}}=\omega /\beta $, the permittivity elements are simplified as

\begin{gather}
\varepsilon _{\text{a,nl}}(k,\omega )=1-\chi \frac{\omega _{p}^{2}}{\omega
^{2}}\text{ , }\varepsilon _{\text{t,nl}}(k,\omega )=1\mathbf{-}\chi \frac{%
\omega _{p}^{2}}{\omega ^{2}-\chi ^{2}\omega _{c}^{2}}\text{, } \notag \\
\varepsilon _{\text{g,nl}}(k,\omega )=\frac{-1}{\omega }\frac{\chi ^{2}\omega _{c}\omega
_{p}^{2}}{\omega ^{2}-\chi ^{2}\omega _{c}^{2}}.  \label{eps}
\end{gather}%
In the limit of $\beta \rightarrow 0 $, then $\chi \rightarrow 1$ and local Drude permittivity model is recovered.

\section{\label{sec:level4} Bulk modes in the hydrodynamic material model}

\bigskip A plane wave in a gyrotropic medium satisfies the Maxwell's equations%
\begin{equation}
\mathbf{k\times E}=\omega \mu _{0}\mathbf{H}\text{ , \ \ }\mathbf{k\times H}%
=-\omega \varepsilon _{0}\overline{\mathbf{\varepsilon }}_{r}\mathbf{(k,}%
\omega \mathbf{)}\cdot \mathbf{E}  \label{d-1}
\end{equation}
For spatially dispersive materials, the response of a particle at position $%
\mathbf{r}$ depends on what happened to the particle at position $\mathbf{r%
%TCIMACRO{\U{b4}}%
%BeginExpansion
{\acute{}}%
%EndExpansion
}$. In this condition, the displacement vector is given as $\mathbf{D}(%
\mathbf{r,}\omega )=\varepsilon _{0}\int \overline{\mathbf{\varepsilon }}%
_{r}(\mathbf{r,r%
%TCIMACRO{\U{b4}}%
%BeginExpansion
{\acute{}}%
%EndExpansion
,}\omega )\cdot \mathbf{E(r%
%TCIMACRO{\U{b4}}%
%BeginExpansion
{\acute{}}%
%EndExpansion
,}\omega )d^{3}\mathbf{r%
%TCIMACRO{\U{b4}}%
%BeginExpansion
{\acute{}}%
%EndExpansion
}$. In a nonlocal homogeneous medium, $\overline{\mathbf{\varepsilon }}_{r}(%
\mathbf{r,r%
%TCIMACRO{\U{b4}}%
%BeginExpansion
{\acute{}}%
%EndExpansion
,}\omega )=\overline{\mathbf{\varepsilon }}_{r}(\mathbf{r-r%
%TCIMACRO{\U{b4}}%
%BeginExpansion
{\acute{}}%
%EndExpansion
,}\omega )$. Then, using the convolution theorem in space domain and spatial Fourier transform we have $\mathbf{D}(\mathbf{k,}\omega )=\varepsilon _{0}\overline{%
\mathbf{\varepsilon }}_{r}(\mathbf{k,}\omega )\cdot \mathbf{E(k,}\omega )$. The wave equation $\left( k_{0}^{2}%
\overline{\mathbf{\varepsilon }}_{r}(\mathbf{k}, \omega)-k^{2}\mathbf{\bar{I}%
}+\mathbf{kk}\right) \cdot \mathbf{E=0}$ is obtained by combining the Ampere and Faraday equations and using the vector identity $\mathbf{k}\times
\left( \mathbf{k\times A}\right) =\mathbf{kk\cdot A}-k^{2}\mathbf{\bar{I}}%
\cdot \mathbf{A}$. Then, the non-zero solutions of $\mathbf{E}$ exists only if $%
\left\vert k_{0}^{2}\overline{\mathbf{\varepsilon }}_{r}(\mathbf{k}, \omega%
)-k^{2}\mathbf{\bar{I}}+\mathbf{kk}\right\vert =0$. Since we are looking for the bulk modes propagating in a plane perpendicular
to the static magnetic vector, we set $k_{z}=0$ in the above determinant, assuming that the in-plane magnetic bias is along the $%
z $-direction, $\mathbf{B}=B_{0}\widehat{\mathbf{z}}$. For this particular case, the determinant is
simplified to two equations, $k_{\text{TM}}^{2}=k_{0}^{2}\varepsilon _{\text{eff}}$ and $k_{%
\text{TE}}^{2}=k_{0}^{2}\varepsilon _{a,nl}$, where $\varepsilon _{\text{eff}}= \left(
\varepsilon _{t,nl}^{2}-\varepsilon _{g,nl}^{2}\right) /\varepsilon _{t,nl}$ and the permitivity
elements are defined in (\ref{eps}). In the local Drude model, these modes are corresponding to
the nontrivial TM and trivial TE modes. 

Figure \ref{disp} shows the dispersion diagram of the trivial and nontrivial bulk modes for nonlocal hydrodynamic and local Drude models. It displays where the nonlocality has significant effect on the dispersion properties. As shown in Fig. \ref{disp}a, the high frequency bands of both models are completely matched. The left inset plot shows that the low frequency bands are also matched for relatively small wavenumbers. The difference appears at very large wavenumbers according to the log scale inset plot on the right side. Fig. \ref{disp}b demonstrates that the trivial modes of both models are identical for the entire momentum domain.

In the local case, the low frequency band is asymptotic to a constant value. This behavior proposes a thermodynamic paradox, because it suggests infinite energy in a limited frequency range, meaning that at $k\rightarrow\infty$ the plasmonic material is still polarized which is not a physically correct behavior. This problem can be solved by including nonlocality in the material model. As seen, the flat parts of the low frequency band wing up when nonlocality is included in the material model via the hydrodynamic model. It also can be understood by looking at (\ref{eps}), where for $k\rightarrow\infty$, all permittivity elements converge to the high frequency dielectric constant.

Topological surface wave (plasmonic or polaritonic) emerges in two different scenarios:  asymmetry in cut-off, or asymmetry in flat asymptote \cite{monticone2020truly}. If the emergence is due to flat asymptote, including nonlocality largely affects at large momentum values. However, considering spatial dispersive models for topological plasmonic or polaritonic structures with periodicity is not crucial due to the finite Brillouin zone.

By adding a realistic level of loss to the hydrodynamic model, the band dispersion is rather modified, but there are still distinguishable bands in the Voigt configuration. Dissipation might lead to topological phase transition, but the presence of damping does not mimic spatial cut-off in the material response. The role of spatial cut-off is setting a bound such that as $k \rightarrow \infty$, permittivity becomes 1 (lossless vacuum), however the effect of loss at large wavenumber is different. Also, in Ref. \cite{shastri2020dissipation}, it has been derived that in the topological Weyl systems,
topological phase transition to a trivial state occurs when unrealistic large dissipation is considered (the Weyl exceptional rings with opposite charges overlap and neutralize each other).
However, a moderate, or low level of dissipation does not redefine topology. In general, dissipation does not lead to topological behaviors alone, because it breaks time-reversal symmetry but not reciprocity. 

\section{Chern numbers}
\begin{figure*} 
	\includegraphics[width=2.0\columnwidth]{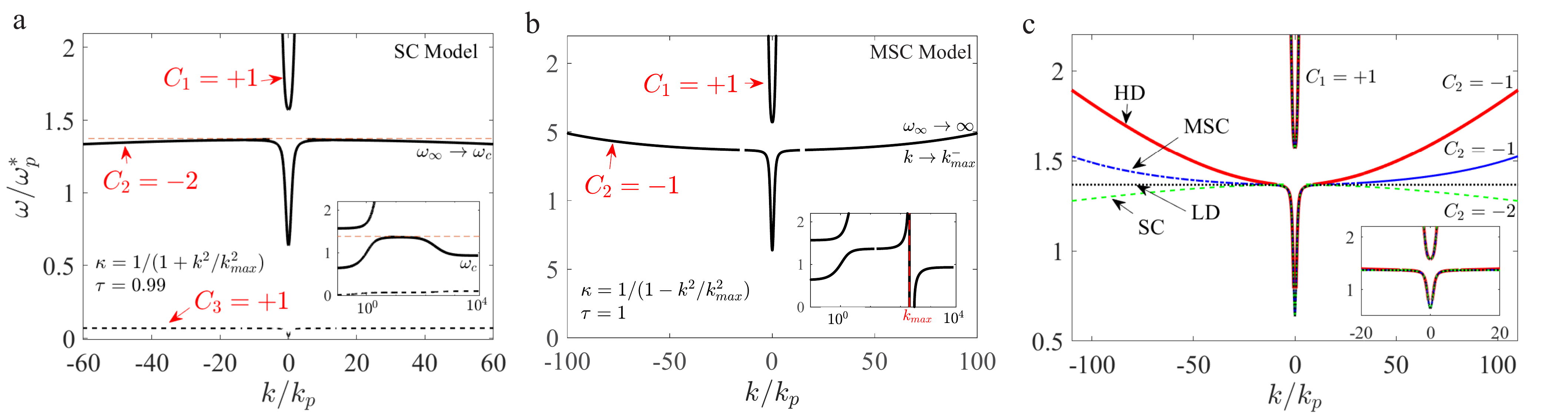}
  \caption{Dispersion bands and associated Chern numbers of (a) spatial cutoff (SC) model with nonlocal factor of $\kappa =%
1/(1+k^{2}/k_{\max }^{2})$, (b) modified spatial cutoff (MSC) model with nonlocal factor $\kappa =%
1/(1-k^{2}/k_{\max }^{2})$, where $k_{\text{max}}=200k_p$. (c) a dispersion plot including dispersion diagram of four material models; SC, MSC, hydrodynamic (HD) and local Drude (LD) models. The magnetized plasma region is characterized by $n_{e}$ $=$ $3.6\times 10^{21}($m$^{-3})$, $%
\varepsilon _{\infty }=15.68$, $m^{\ast }=0.0175$m$_{0}$, $B_{0}=0.6$T, $\mu
=\infty $, given $\omega _{p}^{\ast }=2\pi (1.03$THz$)$, $\omega _{c}/\omega
_{p}=0.23$ and $\gamma =0$.
	}
	 \label{SC}
\end{figure*}

To calculate Chern numbers associated to the frequency bands of the hydrodynamic (HD)
model, we follow the method presented in Ref. \cite{sil2015chern}. For a spatially dispersive material, consider an eigenfunction $%
f_{n}=\left( 
\begin{array}{cc}
\mathbf{E} & \mathbf{H}%
\end{array}%
\right) ^{T}$ with eigenvalues $\omega _{n}$. The envelope of TM\ mode ($%
H_{x}=H_{y}=E_{z}=0$) and TE mode ($E_{x}=E_{y}=H_{z}=0$) are
$f_{nk}^{\text{TM}}=$ $\left( 
\begin{array}{cc}
-\mathbf{\bar{\varepsilon}}^{-1}\cdot \left( \mathbf{k\times \hat{z}}\right)
/\omega _{n}\varepsilon _{0} & \mathbf{1}_{3\times 1}\cdot \mathbf{\hat{z}}%
\end{array}%
\right) ^{T}$ and $f_{nk}^{\text{TE}}=$ $\left( 
\begin{array}{cc}
\mathbf{1}_{3\times 1}\cdot \mathbf{\hat{z}} & \mathbf{k\times \hat{z}/}\mu
_{0}\omega _{n}%
\end{array}%
\right) ^{T}$, respectively. The Berry phase is the phase difference between
the eigenfunctions at $k$ and $k+dk$. It is written in terms of the envelope of electromagnetic field as
\begin{equation}
\mathbf{A}_{nk}=\frac{\text{Re}\{if_{nk}^{\ast }\cdot \frac{1}{2}\frac{%
\partial }{\partial \omega _{n}}(\omega _{n}M)\partial _{\mathbf{k}}f_{nk}\}%
}{f_{nk}^{\ast }\cdot \frac{1}{2}\frac{\partial }{\partial \omega _{n}}%
(\omega _{n}M)f_{nk}}  \label{Ank}
\end{equation}
where $M$ is the material matrix ($M_{11}=\varepsilon _{0}\mathbf{%
\bar{\varepsilon},}$ $M_{22}=\mu _{0}\mathbf{I}_{3\times 3},$ $%
M_{12}=M_{21}=0$). This Berry phase relation was first derived for
periodic photonic crystal structures \cite{raghu2008analogs,haldane2008possible}. However, it can be used for spatially dispersive continuous materials as proved in Ref. \cite{sil2015chern}. Using the Berry phase vector, the Berry
curvature is determined by $\digamma _{\mathbf{k}}=\partial A_{x}/\partial
k_{y}-\partial A_{y}/\partial k_{x}$. The Chern numbers are calculated by the surface integration of the Berry
curvature over the entire momentum space of the wave vector. In analogy to electromagnetic, Berry phase, Berry curvature and Chern number act like magnetic potential vector, magnetic field, and magnetic flux, respectively. In topological materials, the Chern numbers are integer invarients and sum to zero. To obtain integer Chern invariants for continua media, the momentum space must be a close
surface with no boundaries. To realize this condition, the $k_{x}-k_{y}$ plane, which is
the momentum space of the continuous materials, is mapped into the Riemann
sphere, as suggested in \cite{sil2015chern}. In the hydrodynamic model, as shown in Fig. \ref{disp}a the eigenfunctions are
well-behave at large momentum such that the north pole is mapped to the momentum
at $k\rightarrow \infty $ and the Riemann surface becomes a closed surface. Using Stock's
theorem and the fact that the wave functions are not defined at the origin
and infinity, the surface integral in the Chern number relation is written as
two line integrals around the boundary of the surface near the south and north poles. Then,
\begin{equation}
C_{n}=\frac{1}{2\pi }\int\limits_{k=\infty }\mathbf{A}_{n,\mathbf{k}}\cdot
d\mathbf{l}-\frac{1}{2\pi }\int\limits_{k=0^{+}}\mathbf{A}_{n,\mathbf{k}%
}\cdot d\mathbf{l}.
\end{equation}%
Since the system is $\varphi $ independent due to the symmetry about the $z
$-axis, we have $\mathbf{A}_{n,\mathbf{k}}\cdot d\mathbf{l=A}_{n,\varphi
}kd\varphi $. Thus, the Chern number attributed to the n$^{th}$
eigenmode is calculated by $C_{n}=\underset{k\rightarrow \infty }{\lim }(A_{n,\varphi
=0}k)-\underset{k\rightarrow 0^{+}}{\lim }(A_{n,\varphi =0}k)$. Next, we simplify the Berry phase relation (\ref{Ank}) for the nonlocal hydrodynamic
model as 
\begin{equation}
A_{n,\varphi =0}^{TM}k=\frac{\text{Re}\left\{ \frac{ik^{2}}{2(\omega
_{n}\varepsilon _{0})^{2}}\left\{ \left( \left\vert \alpha _{t}\right\vert
^{2}+\left\vert \alpha _{g}\right\vert ^{2}\right) \beta _{g}+2\alpha
_{t}\alpha _{g}\beta _{t}\right\} \right\} }{\frac{k^{2}}{2\left( \omega
_{n}\varepsilon _{0}\right) ^{2}}\left( \left( \left\vert \alpha
_{t}\right\vert ^{2}+\left\vert \alpha _{g}\right\vert ^{2}\right) \beta
_{t}-2\alpha _{t}\alpha _{g}\beta _{g})+\mu _{0}/2\right) },  \label{Ank1}
\end{equation}%
where
\begin{eqnarray}
\alpha _{t} &=&\frac{\varepsilon _{\text{t,nl}}(k,\omega )}{\varepsilon _{%
\text{t,nl}}^{2}(k,\omega )-\varepsilon _{\text{g,nl}}^{2}(k,\omega )} \\
\alpha _{g} &=&-i\frac{\varepsilon _{\text{g,nl}}(k,\omega )}{\varepsilon _{%
\text{t,nl}}^{2}(k,\omega )-\varepsilon _{\text{g,nl}}^{2}(k,\omega )}\\
\beta _{g} &=&\varepsilon _{0}\left( \omega ^{2}+2\beta ^{2}\chi \right) 
\frac{\chi ^{2}2\omega _{c}\omega _{p}^{2}}{\omega \left( \omega ^{2}-\chi
^{2}\omega _{c}^{2}\right) ^{2}} 
\end{eqnarray}
\begin{eqnarray}
\beta _{t} &=&\varepsilon _{0}\left( 1+2(\frac{\beta }{\omega })^{2}\frac{%
\omega _{p}^{2}\chi ^{2}}{\omega ^{2}-\chi ^{2}\omega _{c}^{2}} -\Theta\right),
\end{eqnarray}
where
\begin{eqnarray}
\Theta=\frac{2\chi
^{4}\frac{-2\beta ^{2}}{\omega ^{2}}\omega _{c}^{2}\omega _{p}^{2}-\chi
^{3}\omega _{p}^{2}\omega _{c}^{2}-\omega ^{2}\omega _{p}^{2}\chi }{\left(
\omega ^{2}-\chi ^{2}\omega _{c}^{2}\right) ^{2}}.
\end{eqnarray}
The details of the computation is in the appendix. 
\bigskip In the limit of $k\rightarrow 0$, the nonlocal factor is $\chi
\rightarrow 1$. Zeros of the HD dispersion equation of $k_{TM}^2=\epsilon_{\text{eff}}k_0^2$ are the poles of $\alpha
_{t}$ and $\alpha _{g}$, i.e. at which $\alpha _{t}\rightarrow \infty $ and $\alpha
_{g}$ $\rightarrow \infty $. Then, since $\alpha _{g}/\alpha _{t}=-i\varepsilon
_{g}/\varepsilon _{t}=\mp i$, we have $\underset{k\rightarrow 0\text{ }}{\lim }%
A_{n,\phi =0}k=\pm 1$. As shown in Fig. \ref{disp}a, both frequency bands of the HD model go to infinity ($\omega _{\infty}\rightarrow
\infty$) when $k\rightarrow \infty$. In this limit, $\varepsilon _{g}=0,$ $\varepsilon _{t}=1$ and subsequently $%
\alpha _{t}=1,\alpha _{g}=\beta _{g}=0.$ Therefore, $\underset{k\rightarrow
\infty ,\text{ \ }\omega _{n}\rightarrow \infty }{\lim }A_{n,\phi =0}k=0$.
Finally, the high and low frequency bands of the
nonlocal hydrodynamic model are respectively assigned by the Chern numbers of 
\begin{eqnarray}
C_{1} &=&\underset{k\rightarrow \infty }{\lim }(A_{n,\phi =0}k)-\underset{%
k\rightarrow 0^{+}}{\lim }(A_{n,\phi =0}k)=0-(-1)=1 \\
C_{2} &=&\underset{k\rightarrow \infty }{\lim }(A_{n,\phi =0}k)-\underset{%
k\rightarrow 0^{+}}{\lim }(A_{n,\phi =0}k)=0-1=-1.
\end{eqnarray}
The Chern numbers are integer invarients and sum of them is zero. By
reversing the magnetic bias, the sign of the Chern numbers becomes opposite. For trivial modes, the Chern number is equal to zero. The dispersion bands of the nonlocal HD model are tagged by the relevant Chern numbers in Fig. \ref{disp}. In the following, we compare the results of the HD model with local Drude (LD) and spatial cutoff (SC) model. 

\begin{table*}
\caption {Comparison of material dielectric tensor and Chern numbers of the LD, SC, MSC and HD models}  \label{comp} 
\centering
\begin{tabular}{ |c|c|c|c|c|c| } 
\hline
\multicolumn{2}{|c|}{Material Model} & Local Drude \cite{sil2015chern} & Spatial Cutoff  \cite{sil2015chern} & Modified Spatial Cutoff & Hydrodynamic \\
\hline
\multicolumn{2}{|c|}{Dielectric tensor}& \multicolumn{4}{|c|}{$\mathbf{\bar{\varepsilon}=}\varepsilon _{\text{t}}(%
\mathbf{\bar{I}-\hat{z}\hat{z}})+i\varepsilon _{\text{g}}(\mathbf{\hat{z}}%
\times \mathbf{\bar{I}})+\varepsilon _{\text{a}}\mathbf{\hat{z}\hat{z}}$} \\
\hline
\multirow{3}{6em}{permittivity elements} & $\varepsilon _{\text{a}}(k,\omega )$ & $1-\frac{\omega
_{p}^{2}}{\omega ^{2}}$ & \multicolumn{2}{|c|}{$1-\kappa \frac{%
\omega _{p}^{2}}{\omega ^{2}}\text{ \ }$} & $1-\chi \frac{\omega _{p}^{2}}{\omega ^{2}%
}$\\
 & $\varepsilon _{\text{t}}(k,\omega )$ & $1-\frac{\omega
_{p}^{2}}{\omega ^{2}-\omega _{c}^{2}}$ & \multicolumn{2}{|c|}{$1-\kappa \frac{%
\omega _{p}^{2}}{\omega ^{2}-\omega _{c}^{2}}$} & $1\mathbf{-}\chi \frac{\omega _{p}^{2}}{%
\omega ^{2}-\chi ^{2}\omega _{c}^{2}}$\\
  &$\varepsilon _{\text{g}}(k,\omega )$ & $\frac{\omega _{c}\omega _{p}^{2}}{\omega
\left( \omega _{c}^{2}-\omega ^{2}\right) }$ & \multicolumn{2}{|c|}{$\kappa \frac{\omega _{c}\omega
_{p}^{2}}{\omega \left( \omega _{c}^{2}-\omega ^{2}\right)}$} & $\frac{-1}{\omega }\frac{\chi ^{2}\omega
_{c}\omega _{p}^{2}}{\omega ^{2}-\chi ^{2}\omega _{c}^{2}}$\\
  \hline
\multicolumn{2}{|c|}{nonlocality factor} & - & $\kappa =\frac{1}{1+k^{2}/k_{\max }^{2}}$& $\kappa =\frac{1}{1-k^{2}/k_{\max }^{2}}$&$\chi =%
\frac{1}{1-k^{2}/k_{\max }^{2}},(k_{\max }=\frac{\beta }{\omega })$ \\
\hline
\multicolumn{2}{|c|}{\multirow{2}{*}{Chern numbers}} & $C_{1}=+1$ & $C_{1}=+1$ & $C_{1}=+1$ & $C_{1}=+1$\\
  \multicolumn{2}{|c|}{} & $C_{2}=-1-\text{sgn}(\omega _{_{c}})/\sqrt{1+(\omega _{p}/\omega _{_{c}})^{2}%
}$ & $C_{2}=-2$ &$C_{2}=-1$ &$C_{2}=-1$ \\
\hline
\multicolumn{2}{|c|}{\multirow{4}{*}{Features}} & non integer Chern number,  & integer Chern number,  & integer Chern number, & integer Chern number,\\
\multicolumn{2}{|c|}{} & $C_{2}\notin 
%TCIMACRO{\U{2124} }%
%BeginExpansion
\mathbb{Z}
%EndExpansion
$&$C_{i}\in 
%TCIMACRO{\U{2124} }%
%BeginExpansion
\mathbb{Z}
%EndExpansion
$ & $C_{i}\in 
%TCIMACRO{\U{2124} }%
%BeginExpansion
\mathbb{Z}
%EndExpansion
$ &  $C_{i}\in 
%TCIMACRO{\U{2124} }%
%BeginExpansion
\mathbb{Z}
%EndExpansion
$ \\
 \multicolumn{2}{|c|}{} & non zero sum, $\sum_{i}C_{i}\neq 0$ & non zero sum, $\sum_{i}C_{i}\neq 0$ &zero sum, $\sum_{i}C_{i}=0$ &zero sum, $\sum_{i}C_{i}=0$ \\

\multicolumn{2}{|c|}{} & number of bands:$n_i$=2& $n_i$=2& $n_i$=2&$n_i$=2\\
\hline
\multicolumn{6}{|c|}{Note: For SC model by considering the interpolate material response, the numbers of bands increases to $n_i=3$ with Chern numbers $\{+1,-2,+1\}$  } \\

\multicolumn{6}{|c|}{ and zero sum $\sum_{i}C_{i}=0$, but this method has difficulties of interpolating the interfaced materials. } \\
\hline
\end{tabular}
\end{table*}

For local Drude model case, the Chern number related to the low frequency band is $C_2=-1-\text{sgn}(\omega_{c})%
/(\sqrt{1+(\omega _{p}/\omega _{c})^{2}})$ as determined in \cite{sil2015chern,hanson2016notes}. It is not an integer index, because this band is not well-behaved at large
momentum and it converges to $\ \omega
_{\infty}=\sqrt{\omega _{c}^{2}+\omega _{p}^{2}}$ at the $k\rightarrow \infty $ limit. To solve this issue, in Ref. \cite{sil2015chern} it was suggested to consider a nonlocal material model as $\overline{\mathbf{\varepsilon }}%
_{r}(\omega ,k)=\mathbf{\bar{\varepsilon}}_{R}(\omega )\mathbf{+}\kappa 
\mathbf{\bar{\chi}}_{NR}(\omega )$, in which a nonlocal factor $\kappa =1/(1+k^{2}/k_{\max }^{2})$ with a spatial cutoff wavenumber ($k_{\max }$) is manually added to the material response. The dispersion bands and associated Chern numbers of this spatial cutoff model are shown in Fig. \ref{SC}a. At large momentum $k\rightarrow \infty $, the nonlocal factor goes to zero ($\kappa \rightarrow 0$) and the low frequency band steadily converges to $\omega _{\infty} =\underset{\kappa \rightarrow 0}{\lim }\sqrt{\omega
_{c}^{2}+\kappa \omega _{p}^{2}}=\left\vert \omega _{c}\right\vert$. As a consequence, an integer
Chern index of $-2$ is determined for this band. The high frequency band is still assigned by integer
value of $1$, similar to the local Drude model. Since the sum of Chern numbers is not zero, it has been suggested to consider an interpolated material response as $\overline{\mathbf{\varepsilon }}_{r,\tau }(\omega )=\varepsilon _{\infty }%
\mathbf{+}\tau \left[ \mathbf{\bar{\varepsilon}}_{r}(\omega )-\varepsilon
_{\infty }\right] +(1-\tau )\left[ \mathbf{\bar{\varepsilon}}%
_{Drude}-\varepsilon _{\infty }\right]$ \cite{sil2015chern}. The interpolated material model represents a continuous transition between
an isotropic plasma ($\tau \rightarrow 0^{+}$) with the plasma frequency of $%
\omega _{m}$ and a gyrotropic plasma ($\tau \rightarrow 1^{-}$) with the
plasma frequency of $\omega _{p}$. By this assumption, a new
frequency band with Chern number of +1 appears in the very low frequency range as shown by a black dashed line in Fig. \ref{SC}a, so that the sum of Chern numbers becomes zero. In summary, for the spatial cutoff model one solves the problem of non-integer Chern numbers, but must introduce an interpolated response to obtain Chern numbers that sum to zero (also introducing a new mode that may not be physically relevant). 

In this work, we take into account the effect of nonlocality sing the hydrodynamic model to obtain integer Chern numbers for the plasma continua. Comparing to the spatial cutoff model, we obtain the Chern index of -1 for the low frequency band of the HD model. That is because the general behavior of the low frequency band of the hydrodynamic model is different than the corresponding band in the spatial cutoff model. Therefore, the sum of Chern numbers
becomes zero without the need for considering the interpolated material
response ($\tau = 1$). 

Although not the focus of this work, regarding the spatial cutoff mode, we also found that a slight modification in the nonlocal factor can also avoid the need for an interpolated response. If the nonlocal factor in the spatial cutoff model is
modified to $\kappa =1/(1-k^{2}/k_{\max }^{2}),$ the low frequency
band of the SC model resembles the corresponding band in the hydrodynamic model (a more physically realistic
nonlocal model) as shown in Fig. \ref{SC}b for the limit $|k|\leq k_{max}^{-}$, assuming $k_{max}$ is a large number (we call this model the Modified Spatial Cutoff model, MSC).  Consequently, the associated Chern value changes from $-2$ to $-1$. Therefore, the sum of Chern numbers becomes zero without considering the interpolated material response ($\tau=1$). In addition, by this sign change, the bands Chern numbers are equal to those in the HD model. 

Figure \ref{SC}c shows the band diagram and associated Chern index of all four models discussed above; local Drude, spatial cutoff, modified spatial cutoff and hydrodynamic models. There is a summary of the above discussion in Table \ref{comp}, including a comparison of the different material models and Chern numbers.   

\section{Conclusion}
In this work, we considered the hydrodynamic Drude model and evaluated the effect of nonlocality on bulk dispersion bands and associated Chern numbers. We determined integer Chern invarients which validates the topological behavior of plasma continua by considering the effect of pressure in the material model. We compared the results with the spatial cutoff model and local Drude model. We discussed that the hydrodynamic model 
resolved the non integer Chern invarients in the local Drude model, also  nonphysical response at large wavenumbers, and the need to interpolate the interfaced materials permittivity functions in the spatial cutoff model. It provides a physically well-grounded theory.

\section{Appendix}

\bigskip To simplify the Berry phase relation (\ref{Ank}) for the hydrodynamic Drude model, we calculate each term individually. Then we have 
\begin{equation}
\mathbf{\mathbf{\bar{\varepsilon}}^{-1}}=\left[ 
\begin{array}{ccc}
\alpha _{t} & \alpha _{g} & 0 \\ 
-\alpha _{g} & \alpha _{t} & 0 \\ 
0 & 0 & \alpha _{a}%
\end{array}%
\right] \allowbreak 
\end{equation}%
where 
\begin{equation}
\alpha _{t}=\frac{\varepsilon _{t}}{\varepsilon _{t}^{2}-\varepsilon _{g}^{2}%
},\alpha _{g}=-i\frac{\varepsilon _{g}}{\varepsilon _{t}^{2}-\varepsilon
_{g}^{2}}
\end{equation}%
By expanding $\mathbf{\bar{\varepsilon}}^{-1}\cdot \mathbf{k\times \hat{%
z}}/\omega \varepsilon _{0}$, the TM eigenmode envelope is
\begin{equation}
f_{nk}^{TM}=\left( 
\begin{array}{cccccc}
f_{11} & f_{21} & 0 & 0 & 0 & 1%
\end{array}%
\right) ^{T}
\end{equation}%
where

\begin{equation}
f_{11} =\frac{1}{\omega \varepsilon _{0}}(\alpha _{t}k_{y}-\alpha
_{g}k_{x}), \ \ f_{21} =\frac{1}{\omega \varepsilon _{0}}(-\alpha _{g}k_{y}-\alpha
_{t}k_{x})
\end{equation}
In nonlocal case, $\alpha _{t}$ and $\alpha _{g}$ are $k$-dependent, then
using $\partial _{\mathbf{k}}=\frac{\partial }{\partial k_{x}}\mathbf{\hat{x}%
}+\frac{\partial }{\partial k_{y}}\mathbf{\hat{y}}$ we have 
\begin{equation}
\partial _{k}f_{nk}=\left( 
\begin{array}{cccccc}
F_{11} & F_{21} & 0 & 0 & 0 & 0%
\end{array}%
\right) ^{T}
\end{equation}%
where
\begin{eqnarray}
F_{11}=\partial _{\mathbf{k}}f_{11}=\frac{1}{\omega \varepsilon _{0}}\left[
\left( \frac{\varpi _{t}\delta }{k}k_{x}k_{y}-\frac{\varpi _{g}\delta }{k}%
k_{x}^{2}-\alpha _{g}\right) \mathbf{\hat{x}}\right. \notag \\
\left.+\left( \alpha _{t}+\frac{%
\varpi _{t}\delta }{k}k_{y}^{2}-\frac{\varpi _{g}\delta }{k}%
k_{x}k_{y}\right) \mathbf{\hat{y}}\right] 
\end{eqnarray}
\begin{eqnarray}
F_{21}=\partial _{k}f_{21}=\frac{1}{\omega \varepsilon _{0}}\left[ \left(-%
\frac{\varpi _{g}\delta }{k}k_{x}k_{y}-\frac{\varpi _{t}\delta }{k}%
k_{x}^{2}-\alpha _{t}\right) \mathbf{\hat{x}} \right. \notag \\ 
+\left. \left( -\alpha _{g}-\frac{%
\varpi _{g}\delta }{k}k_{y}^{2}-\frac{\varpi _{t}\delta }{k}%
k_{x}k_{y}\right) \mathbf{\hat{y}}\right] 
\end{eqnarray}%
where we used the partial derivative terms
\begin{eqnarray}
\frac{\partial \alpha _{t,g}}{\partial k_{x}} &=&\frac{\partial \alpha _{t,g}%
}{\partial \chi }\frac{\partial \chi }{\partial k}\frac{\partial k}{\partial
k_{x}}=\varpi _{t,g}\delta _{k}\frac{k_{x}}{k} \\
\frac{\partial \alpha _{t,g}}{\partial k_{y}} &=&\frac{\partial \alpha _{t,g}%
}{\partial \chi }\frac{\partial \chi }{\partial k}\frac{\partial k}{\partial
k_{y}}=\varpi _{t,g}\delta _{k}\frac{k_{y}}{k}
\end{eqnarray}%
where $k=\sqrt{k_{x}^{2}+k_{y}^{2}}.$ Then,
\begin{equation}
\frac{\partial }{\partial \omega }(\omega M(k,\omega ))=\left(
\begin{array}{cc}
\left( \beta _{ij}\right) _{3\times 3} & \mathbf{0}_{3\times 3} \\ 
\mathbf{0}_{3\times 3} & \mu _{0}\mathbf{I}_{3\times 3}%
\end{array}%
\right). 
\end{equation}%
Where $\beta _{t,g}=\frac{\partial }{\partial \omega }(\omega \varepsilon
_{0}\varepsilon _{t,g}).$ Using $\alpha _{t}^{\ast }=\alpha _{t}$ and $%
\alpha _{g}^{\ast }=-\alpha _{g}$, in relation (\ref{Ank}) the denominator is written as 
\begin{equation}
\begin{aligned}
D &=f_{nk}^{\ast }\cdot \frac{1}{2}\frac{\partial }{\partial \omega }(\omega
M)f_{nk}  \\
&= \frac{k^{2}}{2\left( \omega \varepsilon _{0}\right) ^{2}}\left[ \left(
\left\vert \alpha _{t}\right\vert ^{2}+\left\vert \alpha _{g}\right\vert
^{2}\right) \beta _{t}-2\alpha _{t}\alpha _{g}\beta _{g}\right]+\mu _{0}/2
\end{aligned}
\end{equation}%
and the numerator is simplified as 
\begin{equation}
N=if_{nk}^{\ast }\cdot \frac{1}{2}\frac{\partial }{\partial \omega }(\omega
M)\partial _{\mathbf{k}}f_{nk}=N_{x}\mathbf{\hat{x}}+N_{y}\mathbf{\hat{y}}
\end{equation}%
where
\begin{eqnarray}
N_{x} &=&\frac{i}{(\omega \varepsilon _{0})^{2}}\left[ \beta _{t}(\alpha
_{t}k_{y}-\alpha _{g}k_{x})^{\ast }\left( \frac{\varpi _{t}\delta }{k}%
k_{x}k_{y}-\frac{\varpi _{g}\delta }{k}k_{x}^{2}-\alpha _{g}\right) \right. 
\notag \\
&&+\beta _{g}(\alpha _{t}k_{y}-\alpha _{g}k_{x})^{\ast }\left( -\frac{\varpi
_{g}\delta }{k}k_{x}k_{y}-\frac{\varpi _{t}\delta }{k}k_{x}^{2}-\alpha
_{t}\right)  \notag \\
&&-\beta _{g}(-\alpha _{g}k_{y}-\alpha _{t}k_{x})^{\ast }\left( \frac{\varpi
_{t}\delta }{k}k_{x}k_{y}-\frac{\varpi _{g}\delta }{k}k_{x}^{2}-\alpha
_{g}\right)  \notag \\
&&+\left. \beta _{t}(-\alpha _{g}k_{y}-\alpha _{t}k_{x})^{\ast }\left( -%
\frac{\varpi _{g}\delta }{k}k_{x}k_{y}-\frac{\varpi _{t}\delta }{k}%
k_{x}^{2}-\alpha _{t}\right) \right]
\end{eqnarray}%
and
\begin{eqnarray}
N_{y} &=&\frac{i}{(\omega \varepsilon _{0})^{2}}\left[ \beta _{t}(\alpha
_{t}k_{y}-\alpha _{g}k_{x})^{\ast }\left( \alpha _{t}+\frac{\varpi
_{t}\delta }{k}k_{y}^{2}-\frac{\varpi _{g}\delta }{k}k_{x}k_{y}\right)
\right. \notag \\
&&+\beta _{g}(\alpha _{t}k_{y}-\alpha _{g}k_{x})^{\ast }\left( -\alpha _{g}-%
\frac{\varpi _{g}\delta }{k}k_{y}^{2}-\frac{\varpi _{t}\delta }{k}%
k_{x}k_{y}\right) \notag \\
&&-\beta _{g}(-\alpha _{g}k_{y}-\alpha _{t}k_{x})^{\ast }\left( \alpha _{t}+%
\frac{\varpi _{t}\delta }{k}k_{y}^{2}-\frac{\varpi _{g}\delta }{k}%
k_{x}k_{y}\right) \notag \\
&&\left. +\beta _{t}(-\alpha _{g}k_{y}-\alpha _{t}k_{x})^{\ast }\left(
-\alpha _{g}-\frac{\varpi _{g}\delta }{k}k_{y}^{2}-\frac{\varpi _{t}\delta }{%
k}k_{x}k_{y}\right) \right].
\end{eqnarray}
So the Berry phase is written as 
\begin{equation}
\mathbf{A}_{nk}^{TM}=\frac{\text{Re}\left\{N_{x}\mathbf{\hat{x}+}N_{y}\mathbf{\hat{y}}\right\}}{D}.
\end{equation}
Then, using $\hat{\phi}=-\sin (\varphi )\mathbf{\hat{x}}+\cos (\varphi )%
\mathbf{\hat{y}}$ we have
\begin{equation}
A_{n\varphi }^{TM}=\mathbf{A}_{nk}^{TM}\cdot \widehat{\phi }=\frac{\text{Re}%
\{-N_{x}\sin \mathbf{(}\varphi \mathbf{)+}N_{y}\cos (\varphi )\}}{D}.
\end{equation}%
At $\varphi =0$ , $k_{y}=0$ and $k=k_{x}$. So
\begin{equation}
A_{n}^{TM}(\varphi =0)=\frac{\text{Re}\{N_{y}(\varphi =0)\}}{D}.
\end{equation}%
Finally, 
\begin{equation}
A_{n}^{TM}(\varphi =0)k=\frac{\text{Re}\left\{ \frac{ik^{2}}{2(\omega
\varepsilon _{0})^{2}}\left\{ \left( \left\vert \alpha _{t}\right\vert
^{2}+\left\vert \alpha _{g}\right\vert ^{2}\right) \beta _{g}+2\alpha
_{t}\alpha _{g}\beta _{t}\right\} \right\} }{\frac{k^{2}}{2\left( \omega
\varepsilon _{0}\right) ^{2}}\left( \left( \left\vert \alpha _{t}\right\vert
^{2}+\left\vert \alpha _{g}\right\vert ^{2}\right) \beta _{t}-2\alpha
_{t}\alpha _{g}\beta _{g})+\mu _{0}/2\right) }.  \label{Ank1}
\end{equation}
This relation can be used for any nonlocal model by defining relevant $\alpha_{t,g}$ and $\beta_{t,g}$ quantities.

\section*{Acknowledgement}

Funding for this research was provided by the National Science Foundation
under grant number EFMA-1741673.

\maketitle
\bibliographystyle{IEEEtran}
\bibliography{refrences.bib}
\end{document}